\documentclass[]{spie}  

\usepackage{amsmath,amsfonts,amssymb}
\usepackage{graphicx}
\usepackage{aas_macros}
\usepackage{floatrow}
\usepackage[colorlinks=true, allcolors=blue]{hyperref}
\newcommand{\Mod}[1]{\ (\mathrm{mod}\ #1)}
\DeclareMathOperator{\atantwo}{atan2}

\title{The James Webb Space Telescope Aperture Masking Interferometer}

\author[a]{A. Soulain}
\author[b,d]{A. Sivaramakrishnan}
\author[a]{P. Tuthill}
\author[b]{D. Thatte}
\author[b]{K. Volk}
\author[b]{R. Cooper}
\author[c]{L. Albert}
\author[c]{\'E. Artigau}
\author[c]{N. Cook}
\author[c]{R. Doyon}
\author[e]{D. Johnstone}
\author[c]{D. Lafreni\`ere}
\author[b]{A. Martel}
\affil[a]{Sydney Institute for Astronomy (SIfA), School of Physics, The University of Sydney, NSW 2006, Australia}
\affil[b]{Space Telescope Science Institute, 3700 San Martin Drive, Baltimore, MD 21218, USA}
\affil[c]{Institut de recherche sur les exoplan\`etes, Universit\'e de Montr\'eal, Montr\'eal H3C 3J7, Canada}
\affil[d]{Johns Hopkins University Department of Physics and Astronomy, 3400 North Charles, Baltimore, MD 21218}
\affil[e]{NRC Herzberg Astronomy and Astrophysics, 5071 West Saanich Rd, Victoria, BC, V9E 2E7, Canada}
\authorinfo{Further author information: (Send correspondence to A. S.)\\A.S.: E-mail: anthony.soulain@sydney.edu.au}

\begin{document} 
\maketitle

\begin{abstract}
In less than a year, the James Webb Space Telescope (JWST) will inherit the mantle of being the world's pre-eminent infrared observatory. JWST will carry with it an Aperture Masking Interferometer (AMI) as one of the supported operational modes of the Near-InfraRed Imager and Slitless Spectrograph (NIRISS) instrument. Aboard such a powerful platform, the AMI mode will deliver the most advanced and scientifically capable interferometer ever launched into space, exceeding anything that has gone before it by orders of magnitude in sensitivity. Here we present key aspects of the design and commissioning of this facility: data simulations (\texttt{ami\_sim}), the extraction of interferometeric observables using two different approaches (\texttt{IMPLANEIA} and \texttt{AMICAL}), an updated view of AMI's expected performance, and our reference star vetting programs.
\end{abstract}

\keywords{JWST, Aperture-Masking Interferometry, Space Interferometry.}

\section{INTRODUCTION}
\label{sec:intro}  

Non-redundant aperture-masking (NRM) interferometry introduces a mask with well selected holes into the pupil plane of a telescope to enable interferometric capabilities. Major ground-based observatories have successfully implemented NRM interferometry using existing instruments and observing procedures. Landmark discoveries such as dusty disks imaged around young stellar objects, mass-loss shells of evolved stars and the fascinating time-varying spiral plumes surrounding dusty Wolf-Rayet systems have been reported among the 50-odd peer-reviewed papers describing results produced by this technique\cite{1999Natur.398..487T, 2008ApJ...675..698T, 2006ApJ...650L.131L, 2007ApJ...661..496M, 2008ApJ...678L..59I}. The modern era of extreme adaptive optics (XAO) pushes this technique to its next level by dramatically reducing atmospherically-induced wavefront error. The last five years have seen the resurgence of this technique, which is now offered on major observatories around the world---the Very Large Telescope (VLT) with SPHERE and recently VISIR, on Gemini South with GPI, and on the Keck with NIRC2. Recent studies have revealed the full potential of this technique on various astrophysical domains such as planetary systems, protoplanetary disks, brown dwarfs, low-mass stars, and more\cite{2015Natur.527..342S, 2016A&A...588A.117S, 2018MNRAS.480.1006L, 2019A&A...628A.101H, 2019A&A...621A...7W, 2019A&A...622A..96C, 2019AJ....157..249G}.

However, the atmosphere still places limitations on NRM performance. Aperture masking suffers from rapid temporal instabilities due to atmospheric scintillation and transparency variations, as well as from differential atmospheric refraction. Space will provide an exceptionally stable environment that translates to improved performance. Factors that were once secondary will now limit space-based NRM contrast performance. Detector-related effects such as flat-fielding, intra-pixel response (IPR), inter-pixel coupling and charge diffusion, as well as operational considerations such as target acquisition repeatability, guiding jitter and pupil wander, will limit contrasts. Such flagship space missions as JWST has developed thorough knowledge of these diverse noise sources, with decades of modeling as well as plans to characterize them in-orbit. This should result in space-based NRM delivering an order of magnitude better performance than its existing ground-based precursors.

The Aperture Masking Interferometry (AMI\cite{2009SPIE.7440E..0YS}) mode on JWST's NIRISS\cite{2012SPIE.8442E..2RD} instrument will offer three medium bandpass filters (3.8, 4.3 and 4.8\,$\mu m$). AMI promises both high sensitivity and high angular resolution---beyond the Rayleigh criteria---down to $0.5\lambda/D$ (60\,mas at 3.8\,$\mu m$). These filters preserve the non-redundancy of the Fourier space sampling (i.e.: $u-v$ coverage) and provide image plane pixel scales comparable to or finer than Nyquist-sampled. 

\begin{figure}[htbp!]
	\centering
	\includegraphics[width=.65\textwidth]{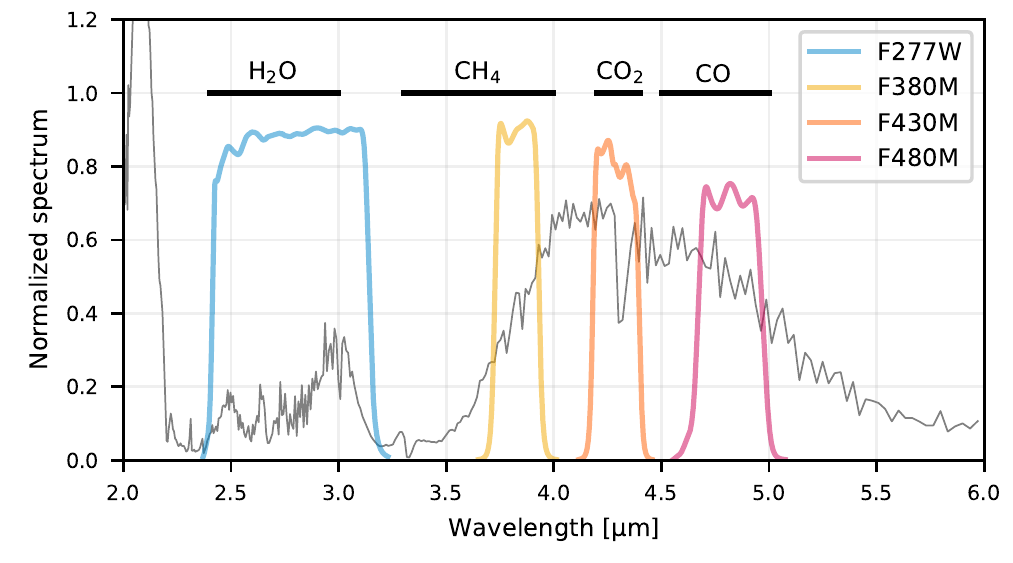} 
	\caption[]{\label{fig:spectra_phillips_niriss}Theoretical spectrum of a cool T-Y brown dwarf\cite{2020A&A...637A..38P} ($\mathrm{T_{eff}}=900$ K, log(g) = 4.5). The NIRISS filters bandwidth available in AMI mode are overplotted. The relevant spectral features of CH$_4$, CO$_2$, CO and H$_2$0 are represented.}
\end{figure}

\begin{figure}[!htbp]
	\centering
	\includegraphics[width=.75\textwidth]{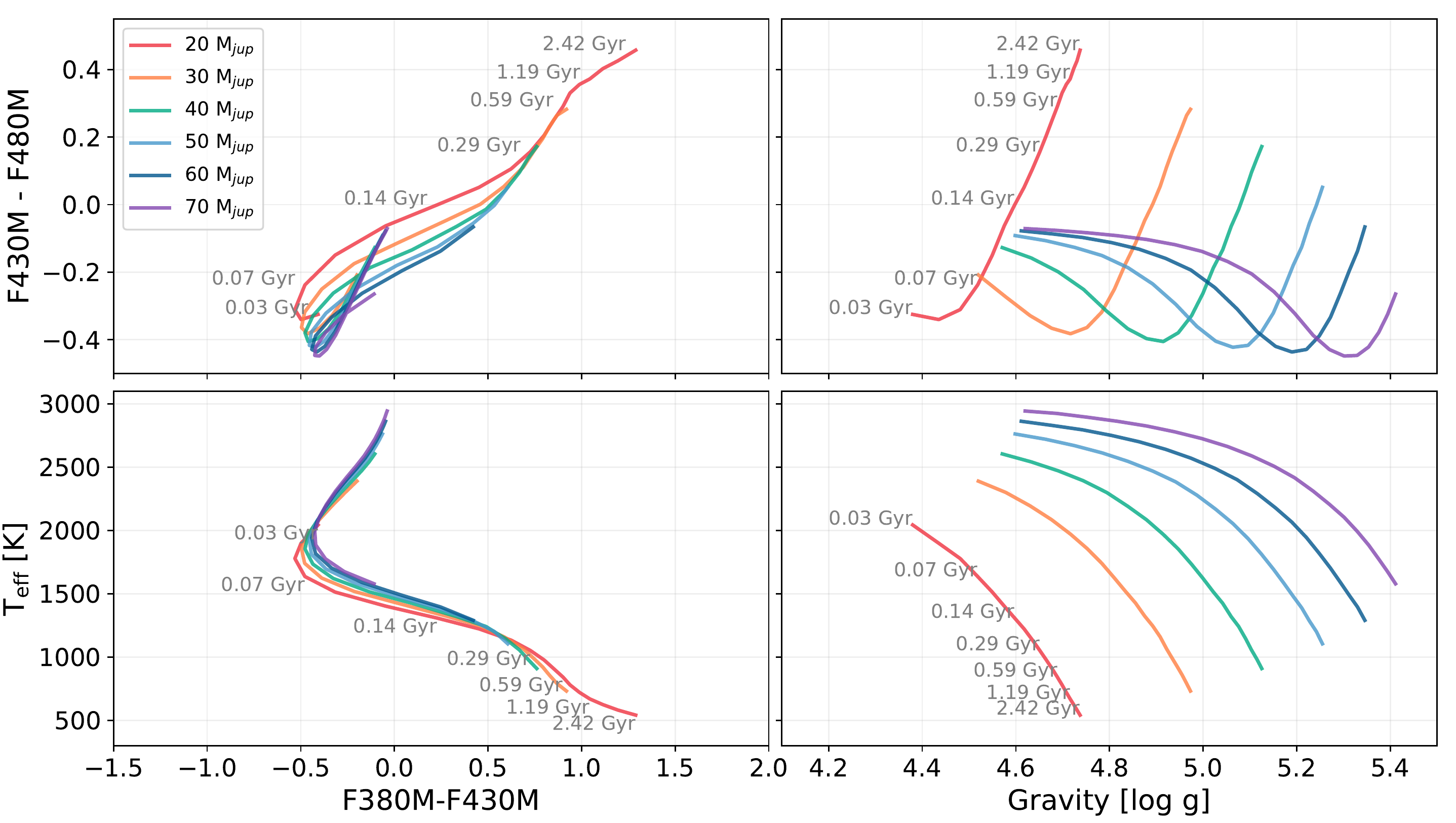} 
	\caption[]{\label{fig:color-plot} Color-color plot (top-left) in AMI bandpasses for brown dwarfs ranging in mass between 20 and 70\,M$_{\rm Jup}$, and ages of 30\,Myr to 2.4\,Gyr. Curves of a given color correspond in all plots to a given mass, while ages are given for the 20\,$_{\rm Jup}$ tracks.  For objects colder than $\sim$1700\,K, the F380M-F430M color provides an excellent proxy of temperature (bottom-left) as it probes the onset of CH$_4$ absorption, a hallmark feature of colder brown dwarfs. The F430M-F480M color is affected by both temperature and surface gravity (top-right), informing on the radius and mass of an object of a given age.}
\end{figure}

 A faint companion detected at sub-Nyquist distance from a host star cannot be observed with NIRSpec or MIRI, and the 3 AMI filters will therefore provide a unique means to constrain its properties, hence the importance of its filter choice. The 3 main AMI filters (Fig.~\ref{fig:spectra_phillips_niriss}) sample the brighter part of the spectral-energy distribution of sub-stellar objects between very strong water features at $<3.5$\,$\mu$m and $>5$\,$\mu$m. The set of filters provides the best constraints on the properties of ultra-cool companions to young stars, being particularly sensitive to the CH4$_2$, CO$_2$ and CO features (3.8, 4.3, 4.8 $\mu m$ respectively \cite{2014SPIE.9143E..40A}). A fourth filter, operating at 2.77 $\mu m$, samples water absorption features although its performance is degraded since NIRISS' pixel scale is coarser than Nyquist at that wavelength. Fig.~\ref{fig:color-plot} shows the evolution of objects in the AMI color-color diagram, as a function of companion age temperature and surface gravity and illustrates how brown dwarfs gradually move through the color-color diagram depending on their bulk properties.

The primary advantage of the AMI mode is its ability to probe the very core of the point spread function (PSF), where coronagraphic techniques on JWST are blind. Thus AMI complements other high contrast JWST imaging on NIRCAM and MIRI\cite{2010PASP..122..162B} in the range of inner working angles between 0.5$\lambda$/D and 4$\lambda$/D which are masked by the JWST coronagraphic spots. This opens the search for companions down to roughly 1~AU for the closest star systems ($\leq50$~pc).

The main scientific application of AMI is for high-contrast detection of point sources, suitable for exoplanets and brown dwarf characterization. With a contrast limit of $10^{-4}$ (Sect. \ref{sec:limits}) within the Rayleigh criterion ($\lambda/D$), AMI will be able to detect a few Jupiter mass exoplanets around nearby young (1-100\,Myrs) stars. For comparison, the current facilities offering such masking modes are limited to a contrast of $10^{-3}$, restraining the scientific cases to the faint binaries\cite{2019A&A...622A..96C, 2019AJ....157..249G}. AMI has also shown its ability to probe the inner structure of nearby Active Galactic Nuclei (AGN), thanks to its relatively good Fourier (or $u-v$) plane coverage\cite{2012SPIE.8442E..2SS}. AMI can benefit from the JWST orientation between two visit to extend its Fourier coverage, filling gaps in the Fourier plane. Such $u-v$ coverage allows image reconstruction using the existing algorithms (\texttt{IRBIS}, \texttt{Squeeze}, \texttt{Mira}, etc.) to resolve the core of circumstellar environments (protoplanetary disks, Wolf-Rayet stars, arc structures, etc.). 

AMI will be particularly suited for follow-up observations of the major giant planet imagers such as SPHERE on the Very Large Telescope (VLT) and GPI on Gemini, providing a complementary wavelength coverage and better overall sampling of their spectral energy distribution (SED) and hence effective temperature and cooling history. The number of exoplanets directly imaged by such instruments continues to rise (51 planets in 2020), indicating a rich field for NIRISS' AMI mode to exploit.

The key advantage of space-based aperture masking will be the dramatically more stable optical system compared to any on the ground. In particular, atmospheric instability typically makes ground-based visibility amplitude calibration laborious, and often scientifically unusable. Space-based observations will add this new dimension to AMI data reduction, and also yield more accurate closure phases than are obtainable from the ground. NIRISS' AMI will deliver interferometric observables that have no ground-based peer.

In this work, we present the advanced simulation tool \texttt{ami\_sim} (Sect. \ref{sec:ami-sim}). We review the two existing methods of extracting interferometric observables (Sect. \ref{sec:extraction}) from image plane interferograms. We apply these tools in an end-to-end analysis to quantify the expected contrast limits of the first space-based aperture mask data (Sect. \ref{sec:limits}). We discuss our candidate calibrator star vetting program which was deemed essential to support AMI's expected performance improvement and present our ground-based measurements of ten calibrators using both adaptive optics instrument and long baseline interferometry (Sect. \ref{sec:vetting}).

\section{ADVANCED SIMULATION TOOLS}
\label{sec:ami-sim} 

In preparation for the start of JWST operations, planned for six months after launch, a few different simulation packages were developed to provide data sets compatible with that of the future facility. \texttt{MIRAGE}\footnote{Available on \url{https://mirage-data-simulator.readthedocs.io/en/latest/}.} is the STscI-supported data simulation tool. It produces raw data cubes (including instrument artifacts), and is used to test STScI's Data Management System (DMS), i.e. JWST's supported data pipeline. But \texttt{MIRAGE} simulations are limited to native pixel size sampling which currently prevents seeding close binaries in a robust manner. In this work, we use a more tailored software,  \texttt{ami\_sim}\footnote{Available on \url{https://github.com/anand0xff/ami_sim}.}, which is dedicated to the simulation of the aperture masking mode of NIRISS. The main advantage of  \texttt{ami\_sim} is its ability to provide bad-pixel-free and reduced (\i.e., post-pipeline) data, coupled with a versatile interface, appropriate for rapidly testing different combinations of astronomical scenes (binary, planetary systems, disks, etc.) and different levels of telescope- and instrument-dependent noise. We also quantify and reduce errors in numerically-generated NRM PSFs that are an input to both data simulations. Our \texttt{ami\_sim} runs utilize these improved PSFs.

\subsection{Non-Redundant Mask Image}
\label{sec:newPSF}
Simulated AMI observations use a set of point-spread functions (PSFs), typically produced using the \texttt{WebbPSF} Python package\cite{2012SPIE.8442E..3DP, 2015ascl.soft04007P}. \texttt{WebbPSF} allows details of the filter bandpass, detector, and input spectral source to be specified, and provides considerable flexibility in the choice of telescope wavefront errors used to calculate the PSF. The pupil mask is defined by a numerical input FITS file. \texttt{WebbPSF} utilizes a matrix Fourier transform which enables a user-specified image pixel scale\cite{2007OExpr..1515935S}. However, there were discrepancies between the numerical \texttt{WebbPSF} simulations made with the previous version of the non-redundant mask image and the PSFs calculated analytically using the ImPlaneIA package\cite{2015ApJ...798...68G, 2018ascl.soft08004G}. What caused this inaccuracy is the input NRM pupil mask image \cite{JWST-STScI-005724}. By constructing the pupil mask on a larger array and binning down to the size required, we create a pupil mask image that produces simulated PSFs that agreed more closely with comparable analytical PSFs. 

\begin{figure}[htbp!]
	\centering
		\includegraphics[width=0.85\textwidth]{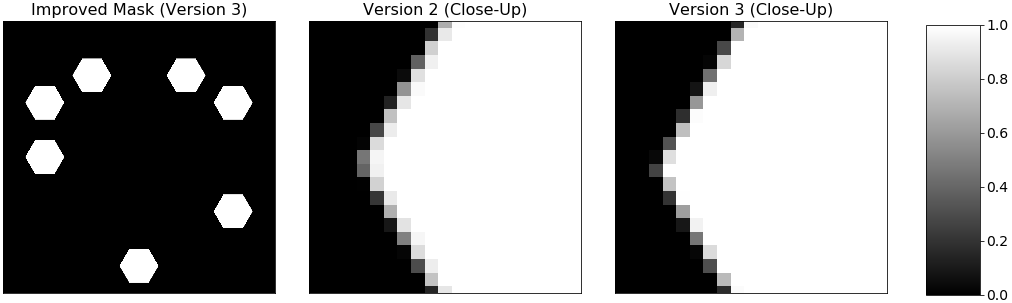} 
	\caption[example] 
	{\label{fig:mask} \textbf{Left:} Improved NRM pupil image. \textbf{Center:} Close-up of the edge pixels of a single sub-aperture in the previous mask version. \textbf{Right:} Close-up of the same region in the improved mask, showing subtle differences in sub-aperture edge pixel values.}
\end{figure}

In order to be able to reliably detect features with contrast as low as $10^{-4}$ in AMI observations, we limited the maximum difference between comparable numerical and analytical PSFs to be of the order of  $10^{-5}$. Using the locations of the centers of the primary mirror segments corresponding to the mask holes and the hexagonal sub-apertures’ flat-to-flat distance of 0.82\,m, we created a square ``parent” pupil mask array of side-length mag~$\times$~1024 (where mag is a positive integer). We binned the parent array down to \texttt{WebbPSF}’s commonly-used 1024$\times$1024 32-bit floating point array. This greyscales or ``anti-aliases” the pupil array we provide to \texttt{WebbPSF}. The geometry of the pupil mask and the difference in the treatment of the pixels at the sub-aperture edges in the improved mask compared to the previous version can be seen in Figure \ref{fig:mask}.

\begin{figure}[htbp!]
	\centering
		\includegraphics[width=0.52\textwidth]{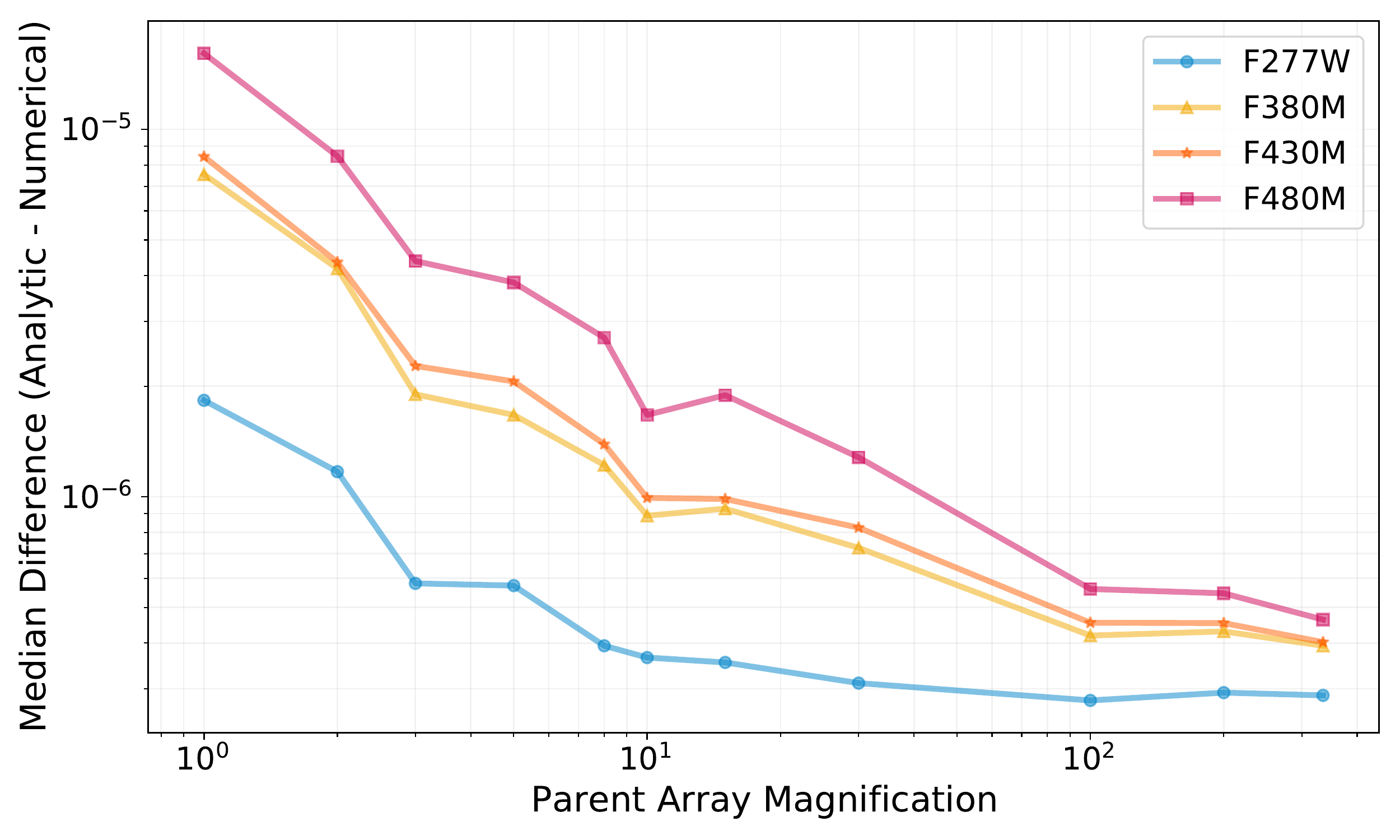} 
	\caption[example] 
	{\label{fig:nrm_improvement} Median difference of analytically computed PSFs minus numerically computed (\texttt{WebbPSF}) PSFs: There is a sharp decrease with increased parent array size until a magnification factor of $\sim100$, remaining approximately constant over the rest of the range of magnifications tested.}
\end{figure}

Agreement between numerical and analytical PSFs improved monotonically with the parent array magnification ``mag”.  At a magnification of 100 we achieved our desired agreement between the two PSFs. The largest parent array we could create (with a 335$\times$ magnification, given the limitations of our computing setup) produced numerical PSFs with maximum differences from the analytical ones on the order of $10^{-5}$ and median differences of the order of $10^{-7}$ between analytical and numerical PSFs when the peak pixel of each PSF is normalized to unity. The trend of median difference with increasing parent array size can be seen in Figure \ref{fig:nrm_improvement}. The improved agreement of the numerical PSFs created using the new mask to the analytical PSFs is supported by the agreement of the central pixel fractions of the numerical PSFs. While the PSFs made using the default pupil mask displayed a 2.8\% difference in the central pixel fraction (CPF) from the analytical ones, the new-mask’s PSF's CPFs differ by 0.17\%.

\subsection{Simulation with \texttt{ami\_sim}}

The new pupil mask presented above was used to create a spatially dependent NRM PSF using \texttt{WebbPSF}\cite{2014SPIE.9143E..3XP}. In turn, that PSF is used by our simulation code (\texttt{ami\_sim}) to generate simulated binary point source image data.

We use a simulation framework that takes two-dimensional input sky scenes from the user and generates simulated observations at a single dither pointing near the center of an 80$\times$80 array with the non-redundant mask (NRM). The dither step is modeled with a 15 mas (rms, single-axis) Gaussian-distributed error. We do not model the pixel response function that is described by Hardy et al., 2014\cite{2014SPIE.9154E..2DH}, or inter-pixel capacitance (IPC) effects. We may implement IPC in a future release. We simulate uncorrelated, normally distributed, 0.1\% standard deviation pixel-to-pixel flat field error, correlated double sampling (CDS) readnoise of 21\,e$^-$, dark current of 0.04\,e$^-$/sec and an assumed background of 0.125\,e$^-$/sec in a pixel.

The AMI mode typically uses the NISRAPID readout pattern, which retains all frames read out up-the-ramp, i.e. one frame per group (NFRAME=1) and no gaps between groups (GROUP\_GAP=0).  This means that after the initial reset that starts the integration, the subarray is continuously read out until the final read out of the integration is followed by a reset. An exposure at each dither position consists of NINT integrations (a.k.a. up-the-ramp) that are placed in a FITS data cube with dimensions FOV$\,\times\,$FOV$\,\times\,$NINT, where FOV is the size of the input target scene (in detector pixels). Each integration is made of NGROUPS up-the-ramp frames, each of TFRAME=0.07544$\,$s duration (which is the frame read time for the SUB80 AMI subarray). The number of groups in an integration (NGROUPS) is chosen to keep the peak pixel below 30$\,$000 e$^{–}$ in the brightest pixel. Integrations are accumulated to reach the required exposure depth (for simplicity we ignore the possibility that we reach the operational maximum number of integrations in the observation in our simulations).

A pointing jitter of 7 mas rms (one-axis) is introduced between each integration. We start with a noiseless count rate input image that is 11 times oversampled and convolve it with the appropriate NRM PSF. The size of the oversampled images is (FOV\,$\times$\,11)\,$\times$\,(FOV\,$\times$\,11) pixels. Thus, we can center our images at 121 different locations (1 pixel/(11$\times$11)) within a detector pixel. For each of the NINT realizations we create a simulated ramp with non-destructively read frames, and add Poisson noise, read noise, dark current, and background realizations to our frames. We then fit a slope to the simulated ramp for each individual pixel using a linear least-squares regression routine and create a ``slope image''. This slope image is divided by a flat field error array to introduce a 0.1\% flat field error standard deviation, and then divided by TFRAME to convert the slope image from counts (\i.e., detected photoelectrons) per frame to counts per second.

When used this way \texttt{ami\_sim}'s output is a datacube ($80$\,$\times$\,80\,$\times$\,$\mathrm{NINT}$).  This corresponds to a JWST post-pipeline dataset, including the different noise sources described above (jitter, flat field, dark current, background, and readnoise). \texttt{ami\_sim} can simulate bad-pixel-free simulated images with realistic noise and flux fluctuation, optimally centered on the detector sub-array. Treatment of missing data from bad pixels prior to extracting interferometric observables is still under development, so we do not generate a realistic bad pixel map in \texttt{ami\_sim}.

\section{EXTRACTING INTERFEROMETRIC OBSERVABLES}
\label{sec:extraction}

There are two different approaches to extracting the relevant quantities (observables) from NRM data: SAMP, which relies on fitting an analytical fringe model in the image plane \cite{2011A&A...532A..72L}, and a second making measurements in the Fourier plane (\texttt{AMICAL}).

\subsection{From the image plane:  \texttt{IMPLANEIA} }
\label{sec:implaneia}

The analytical SAMP fringe fitting methods were extended to include hexagonal sub-aperture shapes in the \texttt{IMPLANEIA} package\cite{2015ApJ...798...68G , 2018ascl.soft08004G}. The advantages of fitting data in the image plane include the ability to ignore bad pixel data, to avoid effects of windowing and tapering in the image plane inherent in numerical Fourier approaches, and to address image plane noise in the image plane itself. \texttt{IMPLANEIA} finds the centroid of the interferogram to sub-pixel accuracy, and generates fringes with a primary beam envelope, which are fit to the data after being binned down to detector pixels.

The outputs include the total flux and the pedestal level of the image, the fringe (or Fourier) phases, and the fringe amplitudes of the interferogram. \texttt{IMPLANEIA} implements a user-selectable image plane oversampling to use when creating its models, and has access to JWST's filter throughput files, as well as STScI’s \texttt{Pysynphot} used to reckon synthetic stellar spectra (e.g.:\, Phoenix model). It accepts three-dimensional cubes of integrations, each of which has the same exposure time, and outputs measurements for each cube in an OIFITS file, following the standard convention to store the interferometric observables\cite{2017A&A...597A...8D}. This data format is broadly use by the interferometric community and compatible with publicly available specialized software developed by JMMC\footnote{Jean-Marie Mariotti Center, \url{http://jmmc.fr}.}.

The package also provides routines to calibrate one OIFITS file with another, to create calibrated OIFITS input data for interferometric image reconstruction or other analysis software.  

\subsection{From the Fourier plane: \texttt{AMICAL}}
\label{sec:amical}

The Aperture Masking Interferometry Calibration and Analysis Library (\texttt{AMICAL}\footnote{Available at \url{https://github.com/SydneyAstrophotonicInstrumentationLab/AMICAL}.}) is a modern Python version of the widely used Sydney aperture masking pipeline developed and improved during the last two decades\cite{1998SPIE.3350..839T, 2000SPIE.4006..491T, 2008ApJ...679..762K}. The code was developed with the main objective to be compatible with existing instruments having an AMI mode. The main advantages of the Fourier sampling approach lie in (1) its flexibility combining different masks, detectors or wavelengths and (2) its speed of execution (few minutes vs. hours).

We focused our efforts to propose a user-friendly interface, though different sub-classes allowing to (1) Clean the reduced datacube from the standard instrument pipelines, (2) Extract the interferometrical quantities (visibilities and closure phases) using a Fourier sampling approach and (3) Calibrate those quantities to remove the instrumental biases.
\\[12pt]
\noindent\textbf{Data processing} The standard calibration procedures followed by the usual pipelines are not sufficient to produce datasets suitable for our interferometric purposes. In the following, we present the extra steps required to calculate the observables.

The initial step consists of cropping the frames to reduce the numerical pressure on the Fast Fourier Transform (FFT), and retrieve a centered image. The centering aspect is essential to avoid numerical artifacts in the Fourier domain. Secondly, several latent issues are caused by the presence of bad pixels. In some cases, the data pipeline does not correct them and they need to be processed carefully during the cleaning step. In \texttt{AMICAL}, we propose a solution to remove these undesirable pixels applying a standard linear interpolation method. The replacement values are computed using the 2D Gaussian kernel convolution available within \texttt{Astropy}\cite{astropy:2018}. In practice, the bad pixel map -- formally retrieved by the JWST pipeline --- is used to generate a list of bad pixel coordinates. These bad pixels are then replaced by the values of the raw image convolved by the 2D Gaussian kernel function at the same native coordinates. This method allows us to interpolate the bad pixels by the values interpolated from a few pixels around, including the same image' noise levels. The use of \texttt{Astropy} package is computationally optimized, allowing to replace a potentially large quantity of pixels in a minimum of time.

Once cleared of bad pixels and centered, we perform background subtraction to ensure a zero pedestal. Then, individual frames are windowed with a super-Gaussian function of the form $e^{-ar^4}$, where $r$ is the radius in pixels and $a=1$ the amplitude of the Gaussian function. The windowing is used to create an interpixel correlation in the Fourier space and ensures to correctly sample the information on each Fourier peaks (so-called splodges). This approach also maintains a smooth transition to zero at the edge of the frames and limits the sensitivity to readout noise. 

\begin{figure}[htbp!]
	\centering
		\includegraphics[width=.42\textwidth]{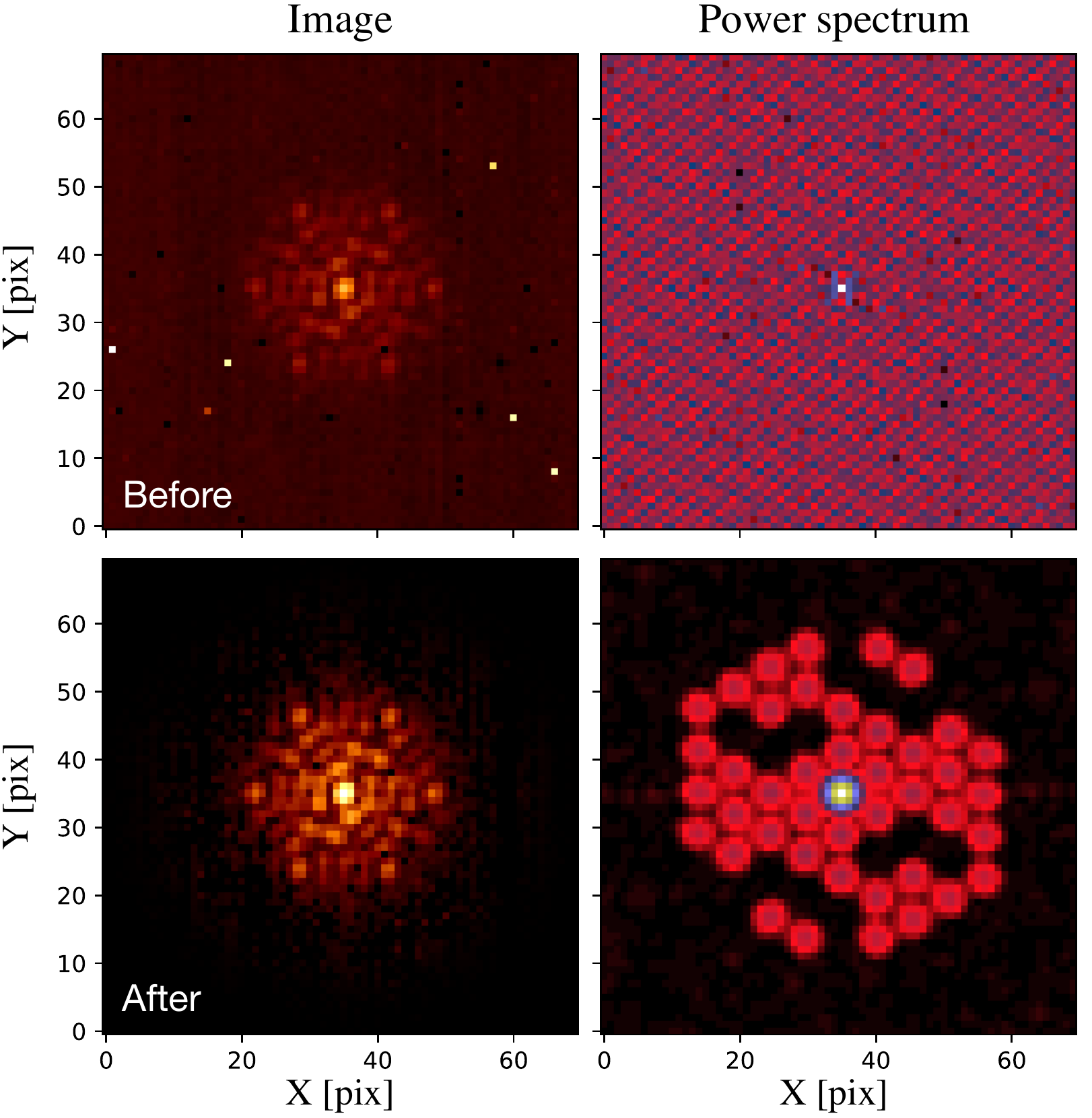} 
	\caption[example] 
	{\label{fig:example_cleaning} Example of reduced post-pipeline dataset from \texttt{MIRAGE} (top) versus cleaned data (bottom) and the associated power spectrum (right).}
\end{figure}

Finally, we perform a first data selection across the cube by maximizing the total flux in each frame (lucky imaging). This selection allows rejecting any bad integrations due to the adaptive optics system, seeing conditions or instrumental issues. In Figure \ref{fig:example_cleaning}, we present an illustration of raw and cleaned subframes and the dramatic impact on the power spectrum.

\paragraph{Extracting the observables} 

\begin{figure}[htbp!]
	\centering
		\includegraphics[width=.75\textwidth]{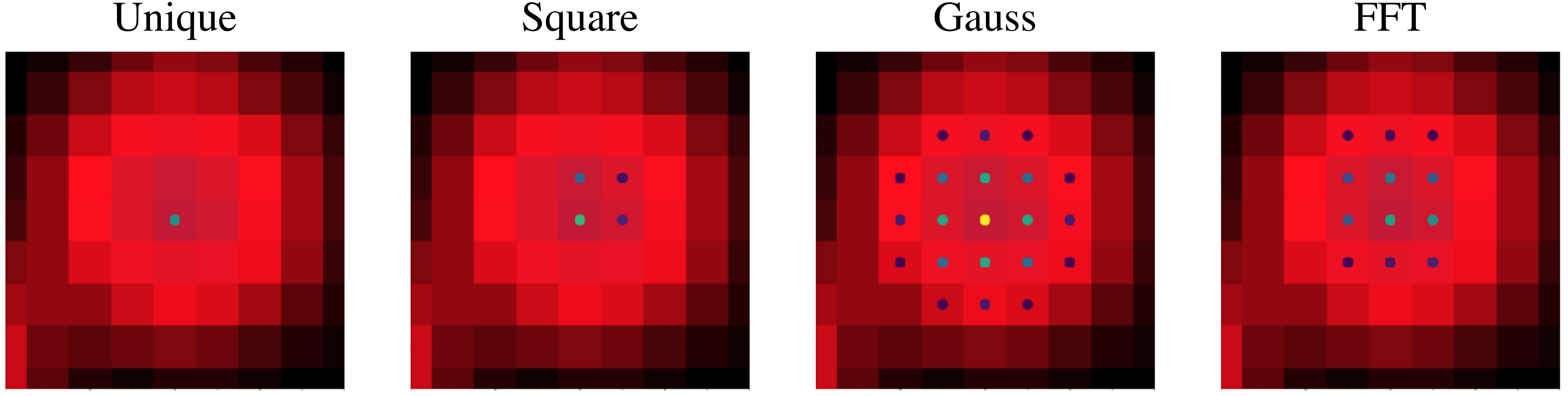} 
	\caption[example] 
	{\label{fig:method_sampling} Different sampling methods implemented in \texttt{AMICAL}. Zoom on one of the 43 splodges (symmetric 21 baselines + central peak, Fig. \ref{fig:example_cleaning}) present in the amplitude map of the Fourier plane.}
\end{figure}

As any interferometric approach, the NRM technique consists of extracting complex quantities encoded within each of the Fourier splodges (Fig. \ref{fig:method_sampling}) where the amplitude of the Fourier peak is identified with the corresponding baseline's fringe visibility (the absolute value of the Fourier transform) and its argument with the fringe phase (imaginary part of the Fourier transform). Although the Fourier phases are corrupted by atmospheric fluctuations and internal optical paths, the observable known as the closure phase (CP) -- the sum of baseline phases around closed triangles -- has been developed by radio astronomers to overcome this problem\cite{1958MNRAS.118..276J}.

Following established practice for NRM data\cite{Tuthill2000a}, we sampled Fourier spectra for each windowed frame depending on the mask coordinates, the wavelength and the pixel size. For a 7-holes mask ($N=7$), the corresponding $u-v$ coverage offers 21 ($N(N-1)/2$) baselines and 35 ($N(N-1)(N-2)/6$) closure phases.

\noindent For each baseline, we compute the $u$ and $v$ Fourier coordinates and convert them to centered detector units with:
\begin{align}
    \label{eq:uvcoord}
    u &= \left(\frac{x_{h1} - x_{h2}}{\lambda} \times \mathrm{FOV}\right) + N_{pix} \Mod{N_{pix}},\\
    v &= \left(\frac{y_{h1} - y_{h2}}{\lambda} \times \mathrm{FOV}\right) + N_{pix} \Mod{N_{pix}},
\end{align}
where $x_{hi}$, $y_{hi}$ are the mask coordinates (m), FOV is the field of view (rad) and $N_{pix}$ the image dimension in detector pixels. 

The finite mask hole size causes information from each baseline to be spread over multiple pixels in the Fourier transform (FT). To exploit information spread beyond just the $u$, $v$ position of the hole centers (Eq. \ref{eq:uvcoord}), we developed four different methods to sample the overall splodges or just a few pixels. The first naive method consists of rounding the $u$, $v$ position on the closest true pixel (``unique'', Fig. \ref{fig:method_sampling}). The second uses the fractional part of the $u$, $v$ position to weight 4 pixels around the expected positions (``square'', Fig. \ref{fig:method_sampling}). The third samples the splodge using a 2D Gaussian weight centered on the rounded $u$, $v$ position (``gauss'', Fig. \ref{fig:method_sampling}). And finally, we apply a ``Fourier optics'' approach using a series of FT applied to the aperture pairs. Then, we threshold and normalize the result to ensure a sum equal to 1 for each splodge (``fft'', Fig. \ref{fig:method_sampling}). This last method takes advantage of computing pixel weights on the ``true'' Fourier grid by the use of FFTs. We recommend this method because of the close-to-Nyquist sampling of the NIRISS detector, which leads to a lack of resolution in Fourier space.

For a given sampling method, we generate the pixel coordinates and the associated weights for each baseline which are used to yield complex visibility data (per frame). 

We then accumulate the robust observables comprising the squared visibilities (V$_{ij}^2$) and closure phases (CP$_{ijk}$) over the data cube. In practice, we compute the squared complex visibility (power spectrum, Eq. \ref{eq:v2}) and the complex triple product (bispectrum, Eq. \ref{eq:bispectrum}) formed by a subset of closing triangles (e.g., apertures 1, 2 and 3). The closure phase is then computed as the argument of this bispectrum (Eq. \ref{eq:cp}). For the V$_{ij}^2$, we measure any bias by taking the average power for regions in Fourier space without any signal. We finally subtract this bias and then normalize by the power at zero baseline (equivalent to normalizing by total flux in the image).
\begin{align}
    \label{eq:v2}
    V^2_{12} &= |C_{12}|^2,\\
    \label{eq:bispectrum}
    BS_{123} &= C_{12}C_{23}C_{13}^*,\\
    \label{eq:cp}
    CP_{123} &= arg(BS_{123}) = \atantwo(BS_{imag}, BS_{real}),
\end{align}
We also implement an alternative method to compute the closure phases using the so-called ``Monnier'' method \cite{1999PhDT........19M}. It consists of averaging many bispectra in each individual image for each triangle of baselines. Once more, the splodges are spread over many pixels which can represent a variety of independent closing triangles. For a given splodge, we define a diameter and account for the different combinations existing to compute the averaged closure phases. This method is computationally more expensive than the Fourier sampling, since hundreds of pixel triangles could exist for each bispectra. However, for most data tested with \texttt{AMICAL}, this method did not present any advantages in terms of scatter or accuracy. We therefore recommend to use the standard sampling approach for NIRISS (or other adaptive optics assisted instruments).

Following the formalism presented by Gordon et al. 2012\cite{2012A&A...541A..46G}, we use their bias-free estimator to compute the closure phase errors (Eq. 40). This method is preferred from the naive standard deviation because it is insensitive to wrapping uncertainty in the complex plane when the variance is large. For the squared visibilities, we used the diagonal of the covariance matrix as a proxy for the errors.

\paragraph{Calibration procedure} Closure phases and square visibilities suffer from systematic terms, caused by the wavefront fluctuations (temporal, polychromatic sources, non-zero size mask, etc.). To calibrate aperture masking data, these quantities are measured on identified point source calibrator stars. In practice, we subtract the calibrator signal from the raw closure phases and normalize the target visibilities by the calibrator's visibilities. 

If several calibrators are available, the calibration factors are computed using a weighted average to account for variations between sources. The extra errors induced are then quadratically added to the calibrated uncertainties. During the calibration procedure, a second data selection is performed to reject bad calibrator-source pairs using a sigma-clipping approach. The calibrated observables are finally stored in \texttt{OIFITS} files.

\section{PLANET AND BINARY DETECTION LIMITS}
\label{sec:limits} 

In this section, we present the updated detection limit estimates achievable with the AMI mode of NIRISS. Previous determinations were performed on cryogenic laboratory data and were focused on the reachable closure phases uncertainties\cite{2015ApJ...798...68G}. In this work, we use the extracted interferometric quantities themselves in an observational approach with the use of standard interferometric tools. Ireland et al. 2013\cite{2013MNRAS.433.1718I} demonstrated that the CP uncertainty is proportional to the number of apertures ($N_{holes}$) and photons ($N_{photons}$) following:
\begin{equation}
    \label{eq:ireland2013}
    \sigma_{CP} = \sqrt{1.5}\times\frac{N_{holes}}{\sqrt{N_{photons}}}
\end{equation}

However, a companion signal in the closure phase is directly proportional to the contrast ratio relative to its host star. For two unresolved stars (or star and planet), the ultimate detection limit can be approximated by the CP uncertainty in radians. According to Eq. \ref{eq:ireland2013}, a contrast of $10^{-4}$ (10 mags) can be reached with $\approx 10^{10}$ photons. This limit represents the goal of AMI-NIRISS and will be investigated using our best understanding of the instrument.

To compute our simulated data set, we used the improved simulated PSF generated based on our revised mask treatment (Sect. \ref{sec:newPSF}). We use \texttt{ami\_sim} (Sect. \ref{sec:ami-sim}) to include the different noise sources (flat-field, background, photon noise, jitter, etc.) and produce the post-pipeline data cube. As for the standard observational procedure, CDS is applied on individual groups (or frames) to reduce them to a single integration, which is accumulated to reach the desired number of photons. The number of groups and integrations are determined using the official Exposure Time Calculator (ETC\footnote{Available on \url{https://jwst.etc.stsci.edu}.}). We considered a sun-like star of spectral star G2V ($\mathrm{T_{eff}=5450\,K}$, $\mathrm{log\,g}=4.5$), represented by the Phoenix spectrum model. We normalize the continuum in the F380M filter of NIRISS to represent a star of magnitude 6 (Vega). This number was chosen to be far enough below the saturation limit, allowing to integrate more groups in a realistic amount of time (Tab. \ref{tab:simu}). Our astrophysical scene is represented by two unresolved components, with a contrast of 20 mags (too faint for NIRISS), a separation of 150 mas and a position angle of 20$^\circ$.

\begin{table}[htbp!]
    \begin{center}
        \caption{Operational parameters used for the AMI-NIRISS simulation (G2V star of 6 mag in F380M filter). The different number of integrations represent the collected number of photons of 10$^7$, 10$^8$, 10$^9$ and 10$^{10}$ respectively.}
        \vspace{.1cm}
        \label{tab:simu}
	    \renewcommand{\arraystretch}{1.3}
		\begin{tabular}{c c c c}
		\hline
		\hline
		& F380M & F430M & F480M\\
		\hline
		Countrate [e$^{-}$/sec] & 1999651 & 1279391 & 1137034\\
		N$_{grp}$ & 7 & 14 & 19\\
		N$_{int}$ & 11, 97, 960, 9590  & 8, 76, 750, 7495 & 7, 63, 622, 6214 \\ \hline
        \end{tabular}
    \end{center}
\end{table}

The interferometric quantities (visibility and CP) are then extracted using the \texttt{AMICAL} software. For NIRISS, we apply the standard procedure consisting of cleaning the data (centering, windowing, background subtraction) and extracting the observables using the FFT method sampling (Sect. \ref{sec:amical}). The contrast limit performances are computed using the CANDID package\cite{2015A&A...579A..68G} between 50 mas ($\approx 0.5\lambda/D$) and 400 mas ($3.5\lambda/D$), which is the inner working angle domain best probed by NIRISS-AMI. Closer companions should be reached using long baseline interferometry or radial velocity techniques whereas outer companions are best observed using coronagraphy techniques. 

\begin{figure}[htbp!]
	\centering
	\floatbox[{\capbeside\thisfloatsetup{capbesideposition={right, top},capbesidewidth=5cm}}]{figure}[\FBwidth]
	{\caption{\label{fig:NIRISSlimits} Expected performance of NIRISS AMI:  3--$\sigma$ contrast as a function exposure depth. The photon noise limited cases are represented as the horizontal dashed-dotted lines (see Eq. \ref{eq:ireland2013}). Colors stand for the number of collected photons and the line styles represent the AMI filters (full, dashed and dotted lines for 3.8 µm, 4.3 µm and 4.8 µm respectively).}}
	{\includegraphics[width=.55\textwidth]{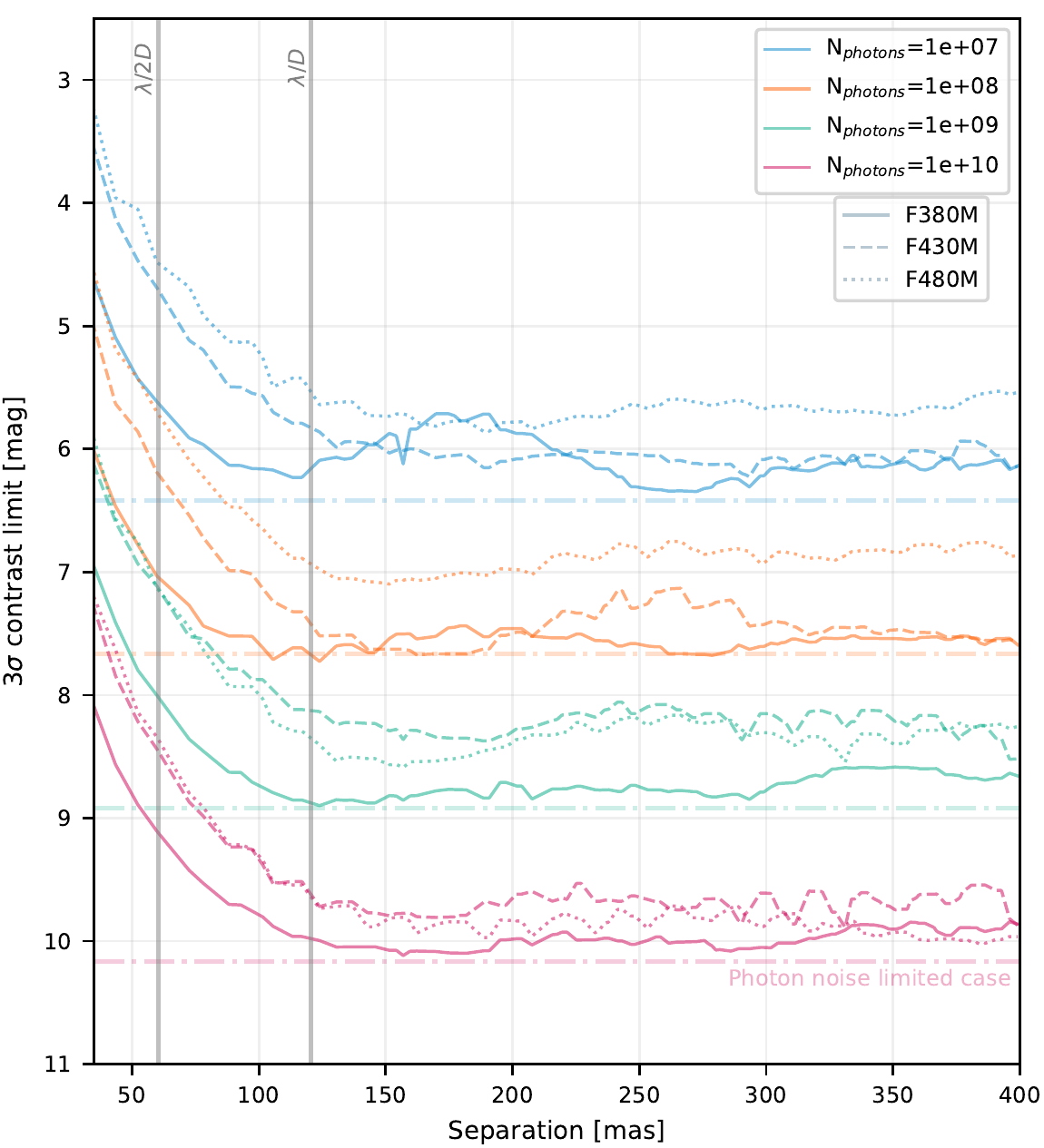}} 
\end{figure}

  We varied exposure depth by simulating data with varying number of photons collected (10$^7$, 10$^8$, 10$^9$ and 10$^{10}$), and derived NIRISS AMI's contrast detection limit curves as a function of separation (Figure \ref{fig:NIRISSlimits}). The given performance correspond to the 3-$\sigma$ confidence interval ($99.73\,\%$) for the three filters of NIRISS (F380M, F430M and F480M). A maximum contrast detection limit of 10 magnitudes is reachable with 10$^{10}$ collected photons (Tab. \ref{tab:limits}), remarkably close to the photon noise limited prediction (Eq. \ref{eq:ireland2013}). We notice that on brighter sources, the total number of photons collectible in a reasonable amount of time ($<$ few hours) and under the technical limitation of 10000 individual integrations per exposure, can be almost doubled allowing to gain up to 0.2 magnitudes in contrast.

\begin{table}[htbp!]
    \begin{center}
        \caption{Contrast performance of AMI-NIRISS between 50 and 400 mas depending on the exposure depths (10$^7$, 10$^8$, 10$^9$ and 10$^{10}$ photons). The given magnitude ranges represent the minimum and maximum values showed in Fig. \ref{fig:NIRISSlimits}.}
        \vspace{.1cm}
        \label{tab:limits}
	    \renewcommand{\arraystretch}{1.3}
		\begin{tabular}{c c c c}
		\hline
		\hline
		N$_{phot}$ & F380M [mag] & F430M [mag] & F480M [mag]\\
		\hline
		10$^{7}$ & 5.4--6.3 & 4.5--6.2 & 4.1--5.9\\
		10$^{8}$ & 6.8--7.7 & 5.9--7.7 & 5.4--7.1\\
		10$^{9}$ & 7.8--8.9 & 6.9--8.5 & 6.7--8.6\\
		10$^{10}$ & 8.9--10.1 & 8.2--9.9 & 8.1--10.0\\
        \end{tabular}
    \end{center}
\end{table}

 The additional source of error ($\approx 15\%$ compared to the theoretical photon noise limited calculation) are probably due a combination of instrumental uncertainties (blurring caused by the jitter, correlated pixel-to-pixel flat field error, etc.) and data processing effects (imperfect Fourier sampling due to the limited field of view, incomplete background subtraction, etc.). We defer the investigation of further sources of noise such as the effects of bad pixel correction, the intra-pixel response (IPR\cite{2014SPIE.9154E..2DH}) or the inter-pixel capacitance (IPC) to a future paper.

\section{CALIBRATOR STAR VETTING PROGRAM}
\label{sec:vetting} 

The vastly improved levels of performance that the NRM interferometric mode of JWST will deliver should contribute significantly in several science topics. With no turbulent atmosphere to disturb the wavefronts, no varying gravity loading to distort the optics, and an exquisitely cold and stable thermal environment, the levels of optical stability provided by the instrument will be unprecedented\cite{2010SPIE.7731E..0FD, 2012SPIE.8442E..2SS}. Nevertheless, in order to benefit from these intrinsic gains in calibration fidelity, it is also necessary to ensure a perfectly known instrumental PSF. To do so, we observed a set of PSF reference stars chosen by the accepted AMI-NIRISS commissioning and observing programs, to verify their suitability. We used among the most advanced ground-base facilities available at the Very Large Telescope (VLT) with SPHERE\cite{2019A&A...631A.155B} and the Very Large Telescope Interferometer (VLTI) with GRAVITY\cite{2017A&A...602A..94G}. 

\subsection{Observation and data reduction}

\subsubsection{Aperture Masking Interferometry with SPHERE}
SPHERE is a high-performance imaging instrument equipped with an Extreme Adaptive Optics system (XAO)\cite{2019A&A...631A.155B}. We observed HD15633, the selected calibrator for the AGN program of NIRISS (PI: K. Ford), and two additional point sources HD15720 and HD16261. Our observations were obtained between October and November 2017 with the IRDIS infrared camera \cite{2008SPIE.7018E..59D, 2014SPIE.9147E..1RL}. We used SPHERE in its interferometric mode using the 7-holes mask available since 2017 \cite{2016SPIE.9907E..2TC}, using the dual-band imaging with K1K2 filter pair (2.110$\pm0.102$ and 2.251$\pm0.109$ $\mu m$).


Data reduction was performed following the procedures described in Zurlo et al. 2016\cite{2016A&A...587A..57Z} with the official ESO-pipeline\footnote{Version 0.4, \url{https://www.eso.org/sci/software/pipelines/index.html}.}. The raw images were reduced by performing background subtraction, bad-pixel correction and flat fielding. We extracted the interferometric observables using the Python package \texttt{AMICAL} (Sect. \ref{sec:amical}). The post-pipeline cubes were cropped, windowed and background subtracted. Once cleaned, we used the FFT method to sample the Fourier space. We calibrated each target by the ones observed closest in time. We finally generated a total of 12 standards calibrated OIFITS2\cite{2017A&A...597A...8D} files for each target.

\subsubsection{Long baseline interferometry with GRAVITY}
The second dataset were observed with VLTI/GRAVITY\cite{2017A&A...602A..94G} in the $K$ band (1.95–2.5$\mu$m) between 2017 and 2018 using the four 1.8\,m Auxiliary Telescopes (ATs). Two targets (HD36805, HD37093) were observed using the small baseline configuration (A0-B2-C1-D0, 11-34m) and five ($\delta$ Crv, HD4981, HD93372, HD93649 and HD101531) using the large configuration (A0-G1-J2-J3, 58-132m). Our sample includes seven different targets, covering four JWST/NIRISS guaranteed time observation proposals (program ID: 1242 (PI: Johnstone), 1200 (PI: Rameau), 1260 (PI: Ford)). HD36805 and HD37093 will be observed during the calibration plan of NIRISS to compute the phase reference. For each target, we observed 2 calibrators (before and after) to retrieve the atmospheric transfer function. 

The data were reduced using the GRAVITY pipeline\cite{2014SPIE.9146E..2DL}, through reduction recipes made available by the
GRAVITY consortium in their python toolkit\footnote{Available at \url{https://version-lesia.obspm.fr/repos/DRS_gravity/python_tools/}.}. Six squared visibilities and four closure phases were obtained with each observation, for both the p and s polarization directions. This yielded two data sets at medium spectral resolution (R=500), and low resolution for the fringe tracker. As there was no significant difference between the two polarisations, we averaged them. We only used the science camera data in our analysis. 

\subsection{Companion search and sensitivity limits}

\begin{figure}[htbp!]
	\centering
		\includegraphics[width=.9\textwidth]{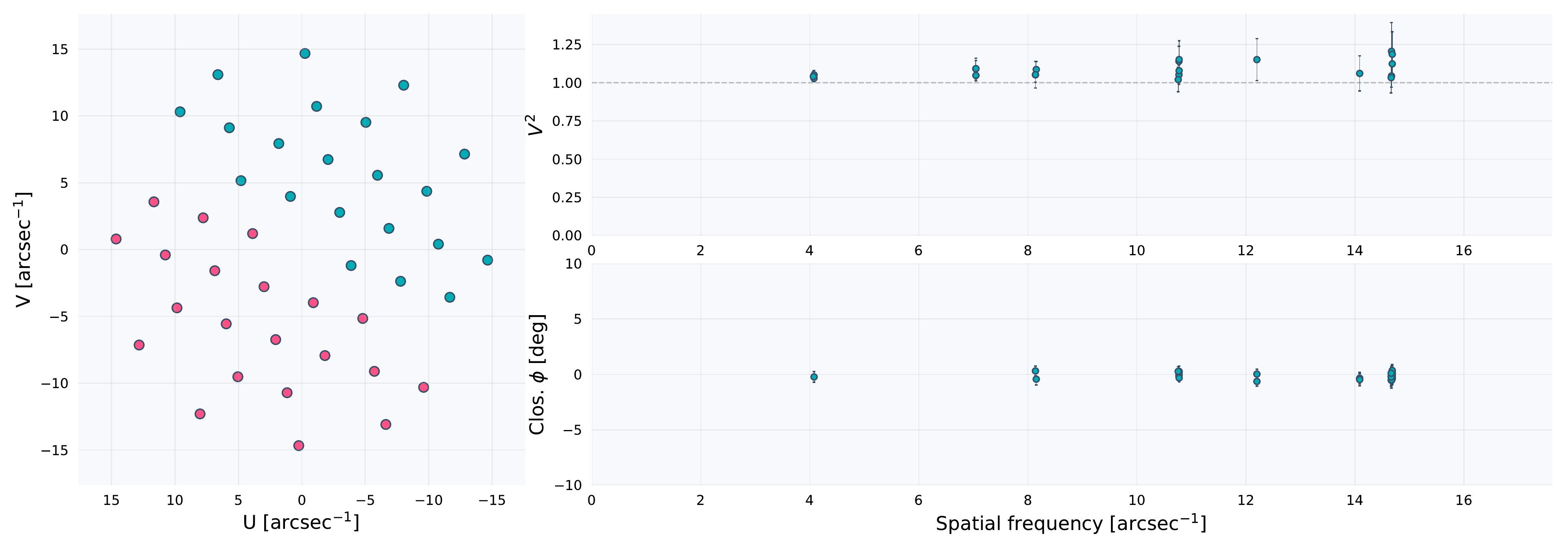} 
	\caption[example] 
	{\label{fig:example_irdis_data} Example of interferometric observables obtained on HD15633. \textbf{Left:} $u-v$ coverage, \textbf{Top-right}: Squared visibilities, \textbf{Bottom-right}: Closure phases. Both visibilities and closure phases are compatible with an unresolved point source (CP $=0$ deg, V$^2=1$).}
\end{figure}

According to our analysis, there is no evidence of multiplicity in our sample. An example of interferometric observables obtained with AMI-SPHERE on HD15633 is presented in Fig. \ref{fig:example_irdis_data} showing both closure phases and squared visibilities compatible with a point source model. In this case, the visibilities seem to be systematically above one (within the uncertainties), a sign of marginally resolved calibrator. 

As for NIRISS, we use \texttt{CANDID}\footnote{Available at \url{https://github.com/amerand/CANDID}} to perform a grid search to probe the immediate surroundings of our reference stars (20-250 mas). At each starting position in the grid, the companion position, its flux ratio, and the angular diameters of the primary are fitted. The \texttt{CANDID} code also includes a tool to estimate the companion detection significance level (in units of sigma). We perform this systematic search using both sets of available observables at once (CP+V$^2$) or separately (CP, V$^2$) without any evidence of multiplicity.

\begin{figure}[htbp!]
	\centering
		\includegraphics[width=.6\textwidth]{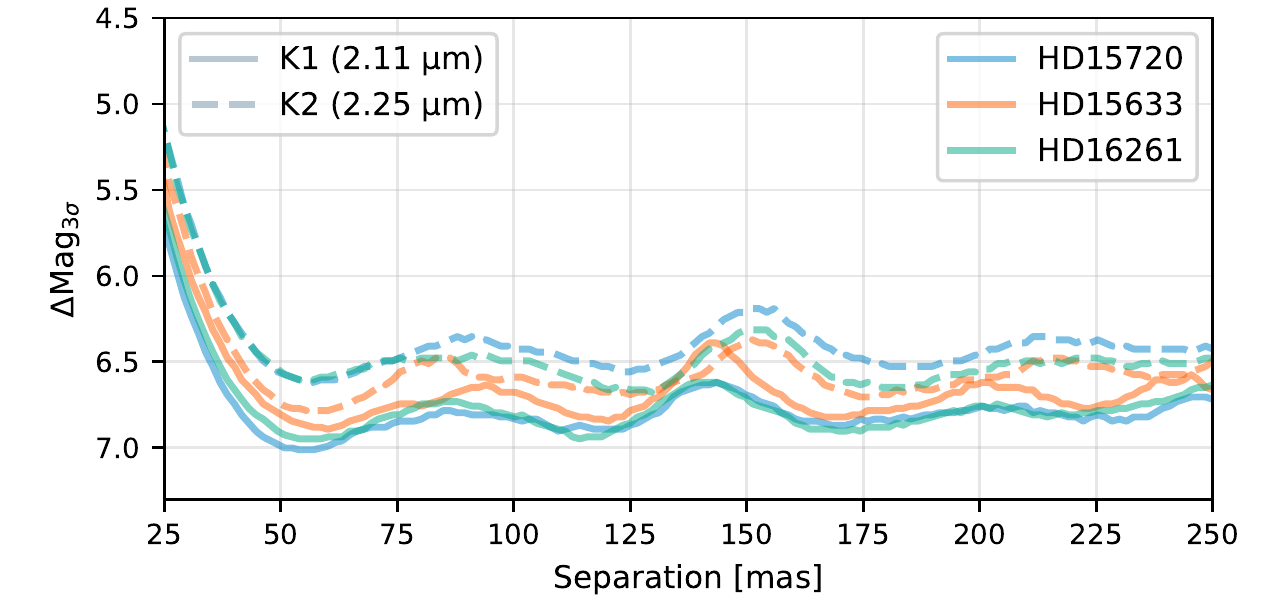} 
	\caption[example] 
	{\label{fig:contrast_limit_sphere}Contrast curves as a function of separation obtained with SPHERE in $K$ band (3--$\sigma$ confidence).}
\end{figure}

For the SPHERE dataset, we utilize \texttt{AMICAL} and the package \texttt{pymask}\footnote{Available at \url{https://github.com/AnthonyCheetham/pymask}.} to compute the sensitivity limits (Fig. \ref{fig:contrast_limit_sphere}). \texttt{pymask} only uses the closure phases which are more reliable measurements and drops the often biased visibilities of the NRM technique. Our best contrast limit was obtained during the first epoch and reached a maximum contrast of 6.8 mags ($\approx520$ in flux ratio) at $\approx\lambda/D$ separation. The contrast ratio achieved with our SPHERE data set is relatively constant between 50 and 250 mas for all calibrators, being between 6.5 to 7 at 2.11 $\mu$m. These contrast limits are notably close to the best contrast ratio reported from the ground with the NRM techniques \cite{2011ApJ...731....8K, 2019JATIS...5a8001S}.

\begin{figure}[htbp!]
	\centering
		\includegraphics[width=.8\textwidth]{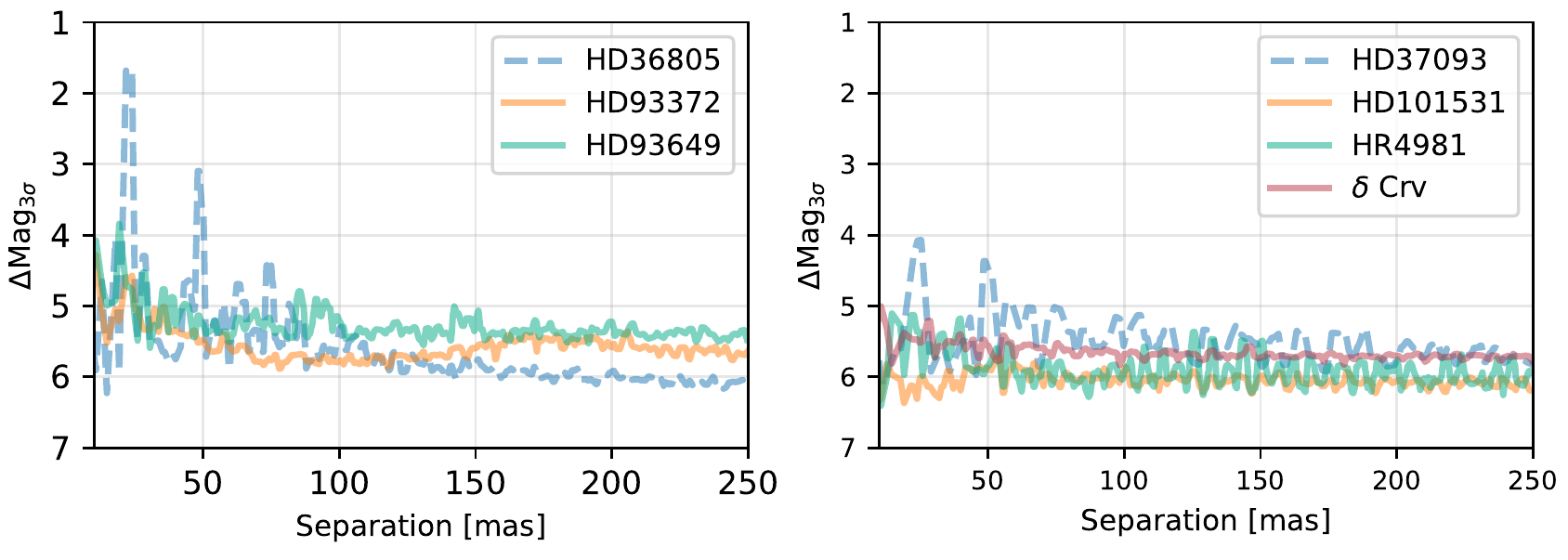} 
	\caption[example] 
	{\label{fig:contrast_gravity} Contrast curves as a function of separation obtained with GRAVITY in K-band (3--$\sigma$ confidence). \textbf{Left:} Limits for the targets using only the visibilities \textbf{Right:} Limits for the targets resolved by GRAVITY (V$^2$+CP).}
\end{figure}

For the GRAVITY sample, the data quality was unequal depending on the baseline configurations and targets. Some targets appeared to be smaller than the associated calibrator and so present visibilities above one. For these stars (HD36805, HD93372, HD93649), we applied the same procedure as for SAM-SPHERE dataset and used only the closure phases to compute their contrast limits (Left, Fig. \ref{fig:contrast_gravity}). 

The other stars ($\delta$ Crv, HD37093, HD101531 and HR4981) were barely resolved by the longest baselines. For these targets, we used \texttt{CANDID} to estimate the contrast limits and include the resolved primary component (Fig. \ref{fig:contrast_gravity}, right panel). The contrast detection limits obtained with GRAVITY are relatively constant across the separation, between 5 and 6 magnitudes (100-250 in flux). We carefully checked the region around the detected peak found for the shortest baseline configuration (around 30 and 50 mas), but without any clear companion detection.

With this study, we confirmed that the 10 reported calibrator stars are trustworthy point sources (in the range of performance of VLT/SPHERE and VLTI/GRAVITY). They are suitable to be observed by the NIRISS consortium during the commissioning and across the GTO programs.

\section{Conclusion}

With the launch planned for October 2021, the James Webb Space Telescope will represent the flagship infrared observatory, surpassing all its precursors in both sensitivity and angular resolution for the next decades. The brand new interferometric mode offered by NIRISS, Aperture Masking Interferometry (AMI),  will push back these limits allowing to probe inside the Rayleigh criterion in an unprecedented way for this spectral-domain (3--5 $\mu$m). As we have seen, the predicted contrast 10$^{-4}$ with AMI-NIRISS is much better than that achieved from the ground (e.g., SPHERE, Fig. \ref{fig:contrast_limit_sphere}). Free of atmospheric constraints, the AMI technique will reach its full potential, opening studies of planetary systems as much as ten magnitudes fainter than their host star. Such contrasts allow characterizing planets of a few Jupiter mass and their parent protoplanetary disk. The promising performance predictions of AMI-NIRISS should inspire further development of masking techniques for the next generation of space telescopes (e.g., NASA/LUVOIR), enabling the full potential of such facilities for a minimal investment.

\acknowledgments

We thanks the JWST/NIRISS consortium both at the IREX institute of Montréal, Canada and at the Space Telescope Science Institute of Baltimore, USA. We acknowledge support from the Australian Research Council (DP 180103408) that funded this work. We also thank the ESO Paranal staff for support for conducting the observations reported in this paper. We thanks the support of the French VLTI center, the ``\textit{Service Utilisateur du VLTI}" (SUV, PI A. Matter OCA-LAGRANGE), especially the help of K. Perraut (IPAG) for the GRAVITY data reduction. We thanks all \texttt{AMICAL} friends for their involvement and work and thus acknowledge C. Robert, A. Cheetham, D. Johnstone, D. Blakely for their assistance. We thanks the major developers involved in the IDL version of the Sydney code J. Monnier, M. Ireland, A. Cheetham among others, and A. Greenbaum,  L. Pueyo, and S. Lacour for their leadership developing analytical approach to extracting observables implemented in \texttt{ami\_sim}.

 \newpage

\bibliography{biblio_ami_spie} 
\bibliographystyle{spiebib} 

\end{document}